%% file: igsc22.tex
\documentclass[conference]{IEEEtran}
\usepackage{cite}

\ifCLASSINFOpdf
  \usepackage[pdftex]{graphicx}
\else
\fi
\usepackage{cite}
\usepackage{amsmath,amssymb,amsfonts}
\usepackage{algorithmic}
\usepackage{graphicx}
\usepackage{textcomp}
\usepackage{xcolor}
\usepackage[a4paper, total={184mm,239mm}]{geometry}
\def\BibTeX{{\rm B\kern-.05em{\sc i\kern-.025em b}\kern-.08em
    T\kern-.1667em\lower.7ex\hbox{E}\kern-.125emX}}

\usepackage{cite}
\usepackage{amsmath,amssymb,amsfonts}
\usepackage{algorithmic}
\usepackage{graphicx}
\usepackage{textcomp}
\usepackage{xcolor}
\usepackage{threeparttable}
\def\BibTeX{{\rm B\kern-.05em{\sc i\kern-.025em b}\kern-.08em
    T\kern-.1667em\lower.7ex\hbox{E}\kern-.125emX}}
\usepackage{multirow}
\usepackage[caption=false]{subfig}
\usepackage{tabu}
\usepackage{color, colortbl}
\newcommand{\thickhline}{%
    \noalign {\ifnum 0=`}\fi \hrule height 1pt
    \futurelet \reserved@a \@xhline
}
\usepackage[linesnumbered,ruled]{algorithm2e}
\usepackage{xcolor}

\SetCommentSty{mycommfont}    
\usepackage{pifont}
\usepackage{cite}
\usepackage{amsmath}
\newtheorem{Tlemma}{Lemma}

\newtheorem{Tdef}{Definition}

\usepackage{amsmath}

\newcommand{\ineq}[1]{\footnotesize$#1$\normalsize}{}
{}

{}
\newcommand{\sm}{\text{{SpiNeMap}}}{}
{}

\begin{document}
\bstctlcite{IEEEexample:BSTcontrol}

%
\title{Design Technology Co-Optimization for Neuromorphic Computing}

\author{\IEEEauthorblockN{Ankita Paul, Shihao Song and Anup Das}
\IEEEauthorblockA{Electrical and\\Computer Engineering\\
Drexel University\\
Philadelphia, PA 19104\\
Email: \{ankita.paul,shihao.song,anup.das\}@drexel.edu}
}


%


\maketitle

\begin{abstract}
\input{sections/abstract}
\end{abstract}


%
\IEEEpeerreviewmaketitle

\section{Introduction}\label{sec:introduction}
\input{sections/introduction}

\section{Background}\label{sec:background}
\input{sections/background}

\section{Design-Technology Tradeoff Analysis}\label{sec:tradeoff}
\input{sections/dtco}

\section{Design Flow Incorporating \\Design-Technology Tradeoff}\label{sec:design_flow}
\input{sections/df}

\section{Conclusion}
\input{sections/conclusions}


\section*{Acknowledgment}
This work is supported by the United States National Science Foundation Faculty Early Career Development Award CCF-1942697 (CAREER: Facilitating Dependable Neuromorphic Computing: Vision, Architecture, and Impact on Programmability) and the United States Department of Energy CAREER Award DE-SC0022014 (Architecting the Hardware-Software Interface for Neuromorphic Computers).



\IEEEtriggeratref{74}
\bibliographystyle{IEEEtran}
\bibliography{commands,disco,external}
%



\end{document}

%% file: sections/abstract.tex
We present a design-technology tradeoff analysis in implementing machine-learning inference on the processing cores of a Non-Volatile Memory (NVM)-based many-core neuromorphic hardware.
Through detailed circuit-level simulations for scaled process technology nodes, we show the negative impact of design scaling on read endurance of NVMs, which directly impacts their inference lifetime.
At a finer granularity, the inference lifetime of a core depends on 1) the resistance state of synaptic weights programmed on the core (design) and 2) the voltage variation inside the core that is introduced by the parasitic components on current paths (technology). We show that such design and technology characteristics can be incorporated in a design flow to significantly improve the inference lifetime.

%% file: sections/introduction.tex
Neuromorphic systems are integrated circuits designed to mimic the computations in a mammalian brain~\cite{mead1990neuromorphic}.
They enable energy-efficient execution of Spiking Neural Networks (SNN)~\cite{maass1997networks} and therefore, these systems are suitable for implementing machine learning inference tasks for embedded Systems and Edge devices in Internet-of-Things (IoT). A neuromorphic system consists of processing cores that implement neurons and synapses. Multiple such cores are interconnected together using Segmented Bus~\cite{balaji2019exploration} or Network-on-Chip (NoC)~\cite{neunoc} to design a many-core neuromorphic hardware~\cite{catthoor2018very}. Table~\ref{tab:hw_examples} illustrates the neuron and synapse capacity of recent neuromorphic hardware cores.

\begin{table}[h!]
	\renewcommand{\arraystretch}{1.2}
	\setlength{\tabcolsep}{1pt}
	\caption{Capacity of recent neuromorphic systems.}
	\label{tab:hw_examples}
	\centering
	\begin{threeparttable}
	{\fontsize{6}{10}\selectfont
		\begin{tabular}{c|cccccccc}
			\hline
			& \textbf{ODIN} & $\mathbf{\mu}$\textbf{Brain} & \textbf{DYNAPs} & \textbf{BrainScaleS} & \textbf{SpiNNaker} & \textbf{Neurogrid} & \textbf{Loihi} & \textbf{TrueNorth}\\
			& \cite{odin} & \cite{mubrain} & \cite{dynapse} & \cite{brainscale} & \cite{spinnaker} & \cite{neurogrid} & \cite{loihi} & \cite{truenorth}\\
			\hline
			\textbf{\# Neurons/core} & 256 & 336 & 256 & 512 & 36K & 65K & 130K & 1M\\
			\textbf{\# Synapses/core} & 64K & 38K & 16K & 128K & 2.8M & 8M & 130M & 256M\\
			\textbf{\# Cores/chip} & 1 & 1 & 1 & 1 & 144 & 128 & 128 & 4096\\
			\hline
			\textbf{\# Chips/board} & 1 & 1 & 4 & 352 & 56 & 16 & 768 & 4096\\
			\hline
			\textbf{\# Neurons} & 256 & 336 & 1K & 4M & 2.5B & 1M & 100M & 4B\\
			\textbf{\# Synapses} & 256 & 336 & 65K & 1B & 200B & 16B & 100B & 1T\\
			\hline
	\end{tabular}}
	\end{threeparttable}
\end{table}

A neuromorphic core can be implemented using an analog crossbar where bitlines and wordlines are organized in a grid with memory cells connected at their crosspoints to store the synaptic weights~\cite{liu2015spiking,hu2014memristor,hu2016dot,ankit2017trannsformer}.
Neuron circuits are implemented along bitlines and wordlines (see Figure~\ref{fig:crossbar}).
Recently, Non-Volatile Memory (NVM) technologies such as phase-change memory (PCM) and oxide-based resistive switching random access memory (OxRRAM) are used to implement the memory cells in a neuromorphic core due to their low power consumption, CMOS-compatible scaling, and multilevel analog operations~\cite{mallik2017design}.\footnote{NVMs are also used in classical von Neumann computing to mitigate the performance and energy bottlenect of DRAM~\cite{palp,mutlu2013memory,datacon,mutlu2015research,mneme,hebe,shihao_igsc}.} In an NVM-based neuromorphic inference hardware, the resistance state corresponding to the synaptic weights of an SNN are programmed on the NVM cells.

Unfortunately, NVMs suffer from reliability issues such as circuit aging, limited endurance, and read disturbance~\cite{shelby2015non,larcher2010high,lu2006non,noe2017phase}. Many of these issues have been addressed recently in the context of neuromorphic computing~\cite{song2020case,balaji2019framework,reneu,vts_das,ncrtm,twisha_endurance,twisha_thermal,espine}. In this work, we focus on the read disturbance issues of OxRRAM technology, where an OxRRAM cell's resistance state may change after performing a certain number of read operations~\cite{shim2020impact,song2021improving,chen2019non}.
Therefore, when an OxRRAM-based neuromorphic hardware is used to implement an inference task, the trained synaptic weights programmed on the hardware may change after performing a few inference operations. To ensure expected results from an inference task, the trained weights need to be reprogrammed on to the hardware periodically.
Reprogramming of synaptic weights to a neuromorphic hardware involves offlining the hardware and transferring the weights from the host to the hardware via a bandwidth-limited interconnect.
This can significantly increase the overhead, and lower the availability and reliability of neuromorphic computing. 

In this work, we present a design-technology tradeoff analysis in implementing machine learning inference on many-core neuromorphic hardware.
Through circuit-level simulations we show the negative impact of design scaling on read endurance of NVMs, which directly impacts their inference lifetime. We show that the inference lifetime of a crossbar depends on 1) the resistance state of synaptic weights programmed on the NVMs and 2) the voltage variation inside the crossbar that is introduced by the parasitic components on current paths. Such design and technology characteristics can be incorporated in a design flow to significantly improve the inference lifetime.

%% file: sections/background.tex
We briefly introduce Spiking Neural Network (SNN) and their implementation in hardware. 

\begin{figure}[h!]
	\centering
	\centerline{\includegraphics[width=0.99\columnwidth]{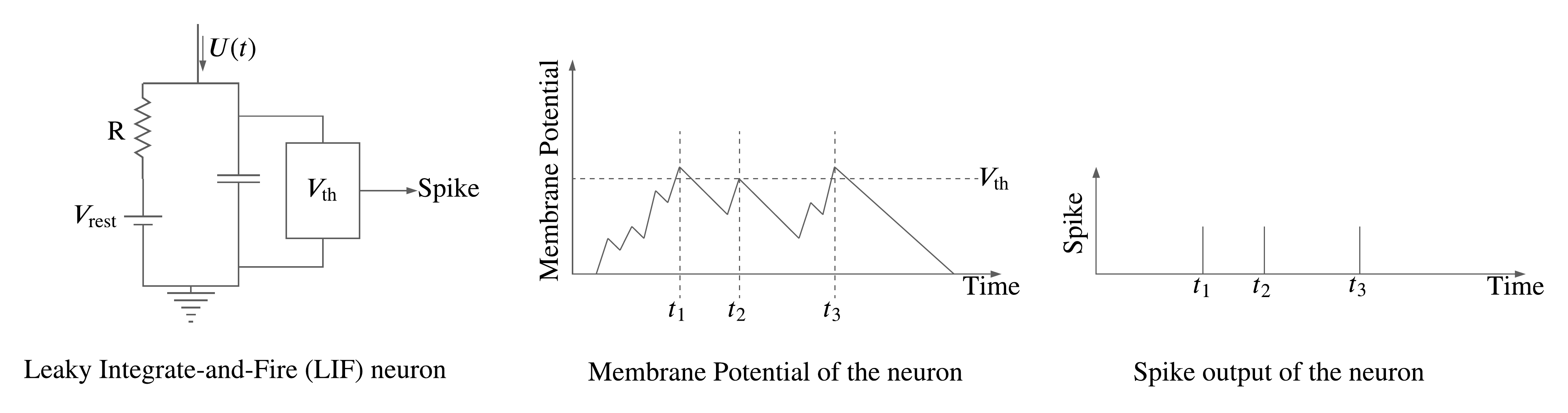}}
	\caption{A leaky integrate-and-fire (LIF) neuron. The membrane potential over time of the neuron (middle). The spike output of the neuron representing its firing time (right).}
	\label{fig:lif}
\end{figure}

Spiking Neural Networks (SNNs) are the third generation of neural networks designed using spiking neurons and bio-inspired learning algorithms~\cite{maass1997networks}. In an SNN, spikes injected from pre-synaptic neurons raise the membrane voltage of a post-synaptic neuron (middle sub-figure of Figure~\ref{fig:lif}). When the membrane voltage crosses a threshold ($V_\text{th}$), the post-synaptic neuron emits spikes that propagate to other neurons (right sub-figure of Figure~\ref{fig:lif}). SNNs implement some variants of Integrate and Fire (I\&F) neurons with a spike duration ranging from 1 $\mu$s to several ms~\cite{ifneuron,zhang2020pulse,leigh2020efficient} (left sub-figure of Figure~\ref{fig:lif}).

SNNs can implement many machine learning approaches~\cite{HeartEstmNN,moyer2020machine,dominguez2018deep,das2018heartbeat,jolpe18,yu2015spiking,schliebs2013evolving,dong2018unsupervised}. In a supervised machine learning, an SNN is pre-trained with representative data. Machine-learning {inference} refers to feeding live data points to this trained SNN to generate the corresponding output.
The {quality} of machine learning inference can be expressed in terms of accuracy~\cite{jolpe18}, Mean Square Error (MSE)~\cite{HeartEstmNN}, Peak Signal-to-Noise Ratio (PSNR)~\cite{carlsim}, and Structural Similarity Index Measure (SSIM)~\cite{hore2010image}.
For SNNs, these quality metrics are defined in terms of the inter-spike interval (ISI)~\cite{cariani2001temporal,fang2020encoding,pycarl,kiselev2016rate}. If  \ineq{\{t_1,t_2,\cdots,t_{K}\}} denote a neuron's firing times in the time interval \ineq{[0,T]}, the average ISI of this spike train is

\begin{equation}
    \label{eq:isi}
    \footnotesize \mathcal{I} = \sum_{i=2}^K (t_i - t_{i-1})/(K-1).
\end{equation}

A neuromorphic hardware platform is implemented as a tiled architecture, where tiles are interconnected via a time-multiplexed shared interconnect.
Each tile consists of a neuromorphic processing core, which can implement neurons and synapses of a machine learning model.
 A common design practice is to use an analog crossbar to implement a core. 
 
\begin{figure}[h!]
	\centering
	\vspace{-5pt}
	\centerline{\includegraphics[width=0.99\columnwidth]{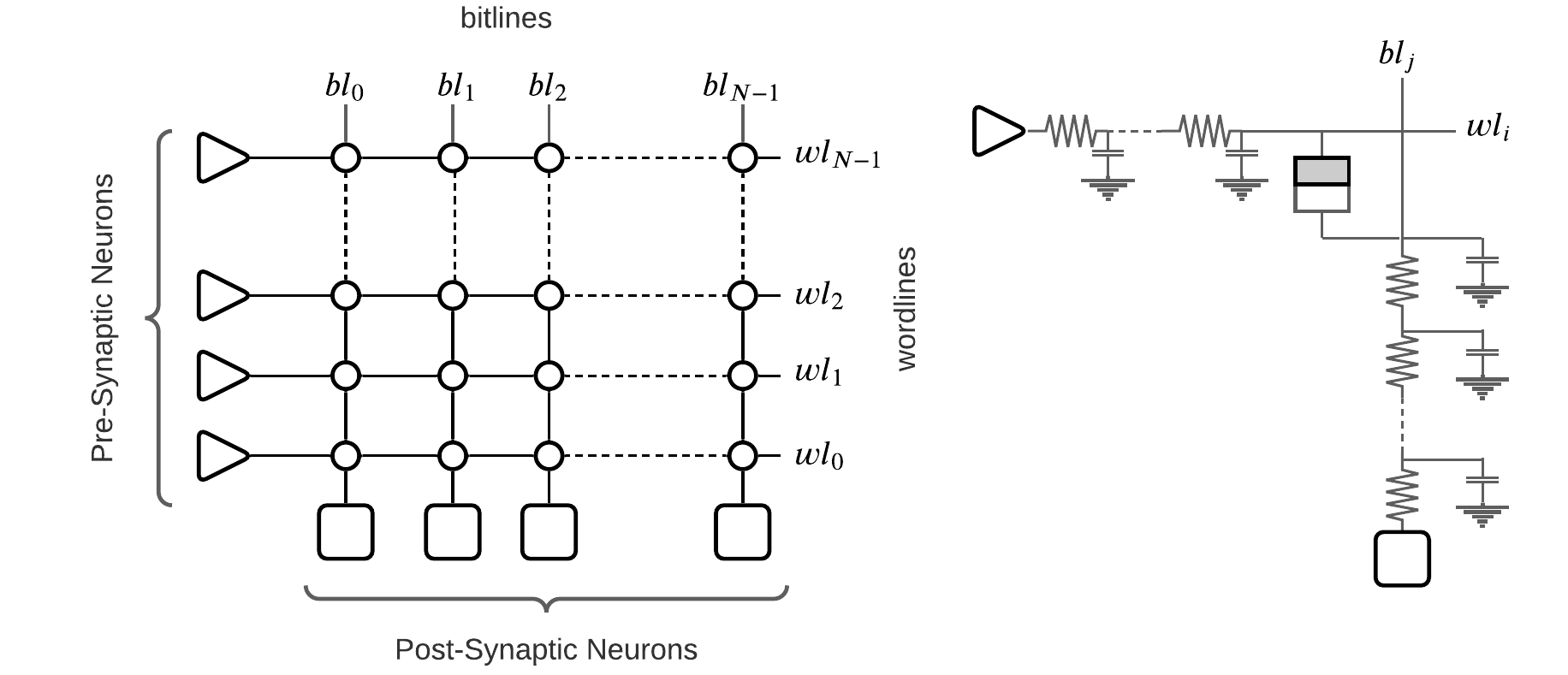}}
	\vspace{-10pt}
	\caption{An $N \times N$ crossbar showing the parasitic components within.}
	\vspace{-5pt}
	\label{fig:crossbar}
\end{figure}

 In a crossbar, pre-synaptic neuron circuits are placed on horizontal wires called wordlines, while post-synaptic neuron circuits are placed on vertical wires called bitlines. Memory cells are placed at the crosspoint of each wordline and bitline, and they implement the synaptic weights of an SNN. The left subfigure of Figure~\ref{fig:crossbar} illustrates an $N \times N$ crossbar. The right subfigure illustrates the parasitic RC components on the current path from a pre-synaptic neuron to a post-synaptic neuron accessing the memory cell $(i,j)$ located at the crosspoint of \ineq{i^\text{th}} wordline and \ineq{j^\text{th}} bitline.

Overall, a neuromorphic hardware enables distributed and pipelined processing of SNN operations. Additionally, each crossbar in the hardware can implement a maximum of \ineq{N} pre-synaptic neurons per post-synaptic neuron. Therefore, system-software frameworks such as NEUTRAMS~\cite{ji2016neutrams}, NeuroXplorer~\cite{neuroxplorer}, Corelet~\cite{corelet}, PACMAN~\cite{pacman}, and LAVA~\cite{loihi_mapping} consist of 1) a compiler, which partitions an SNN model into clusters such that the neurons and synapses of each cluster can be mapped to a crossbar of the hardware, and 2) a run-time manager, which maps the clusters of an SNN to the cores of a many-core hardware. To this end, several mapping strategies have been proposed, including optimizing for
energy~\cite{psopart,spinemap,twisha_energy,balaji2020run}, throughput~\cite{sdfsnn,sdfsnn_pp,dfsynthesizer,dfsynthesizer_pp}, resource utilization~\cite{esl20,adarsha_igsc,loihi_mapping,ji2016neutrams,pycarl}, circuit aging~\cite{ncrtm,balaji2019framework,reneu,song2020case,vts_das}, inference lifetime~\cite{song2021improving}, and write endurance~\cite{twisha_endurance,twisha_thermal,espine}. 
These mapping techniques all use some variant of the SNN-partitioning approach proposed in SpiNeMap~\cite{spinemap}.

Recently, dataflow models have been used to analyze performance of SNNs implemented on neuromorphic hardware, especially using Synchronous Dataflow Graphs (SDFGs)~\cite{lee1987synchronous}.\footnote{SDFGs are commonly used to model streaming applications that are implemented on a multi-core system and their performance analysis~\cite{SB00,jiashu2012design,das2018reliable,das2015reliability,ghamarian2006throughput}.} There are two strategies proposed in literature -- the {SDFSNN}~\cite{sdfsnn} and its extended versions~\cite{sdfsnn_pp,shihao_designflow,shihao_soda}, which uses dataflow graphs to model an SNN, performing partitioning and mapping explorations with neurons and synapses directly, and the {DFSynthesizer}~\cite{dfsynthesizer} and its extended version~\cite{dfsynthesizer_pp}, which uses dataflow graphs to only model the clustered SNN, allowing mapping and scheduling of the clusters (a collection of neurons and synapses) to the PEs of a neuromorphic hardware. 

Emerging non-volatile memory (NVM) technologies such as phase-change memory (PCM), oxide-based memory (OxRAM), spin-based magnetic memory (STT-MRAM), and Flash have recently been used for synaptic storage in crossbars. NVMs are non-volatile, have high CMOS compatibility, and can achieve high integration density. Each NVM device can implement both a single-bit and multi-bit synapse. 
Because of these properties, an NVM-based neuromorphic hardware typically consumes energy that is magnitudes lower than using SRAMs~\cite{Burr2017,kim2015nvm,eryilmaz2015device}. However, NVMs also introduce reliability issues~\cite{shelby2015non}. Table~\ref{tab:reliability_summary} summarizes them for different NVMs.

\begin{table}[h]
\renewcommand{\arraystretch}{1.5}
\setlength{\tabcolsep}{2pt}
\caption{Reliability issues in NVMs.}
\label{tab:reliability_summary}
\centering
{\fontsize{10}{10}\selectfont
\begin{tabular}{|l|c|}
\hline
\textbf{Reliability Issues} & \textbf{NVMs}\\
\hline
High-voltage related circuit aging & PCM, Flash\\
High-current related circuit aging & OxRAM, STT-MRAM\\
Limited read endurance & All\\
Limited write endurance & All\\
\hline
\end{tabular}}
\end{table}

We discuss one specific issue -- limited read endurance for OxRRAM-based neuromorphic hardware.

\section{Introduction to Oxide-based Resistive RAM}

\begin{figure}[h!]
	\begin{center}
		\vspace{-10pt}
		\includegraphics[width=0.69\columnwidth]{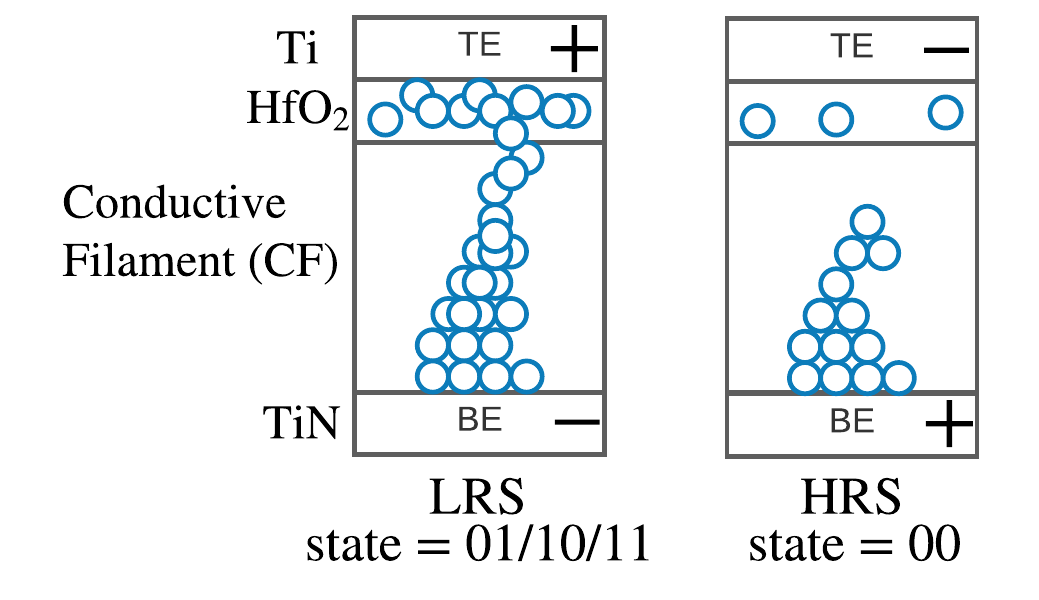}
		\vspace{-10pt}
		\caption{Operation of an OxRRAM cell with the $\text{HfO}_2$ layer sandwiched between the metals Ti (top electrode) and TiN (bottom electrode). The left subfigure shows the formation of LRS states with the formation of conducting filament (CF). This represents logic states 01, 10, and 11. The right subfigure shows the depletion of CF on application of a negative voltage on the TE. This represents the HRS state or logic 00.}
		\label{fig:RRAM}
		\vspace{-10pt}
	\end{center}
\end{figure}

The resistance switching random access memory (OxRRAM) technology presents an attractive option for implementing the synaptic cells of a crossbar due to its demonstrated potential for low-power multi-level operation and high integration density~\cite{mallik2017design}. An OxRRAM cell is composed of an insulating film sandwiched between conducting electrodes forming a metal-insulator-metal (MIM) structure (see Figure~\ref{fig:RRAM}). Recently, filament-based metal-oxide OxRRAM implemented with transition-metal-oxides such as HfO${}_2$, ZrO${}_2$, and TiO${}_2$ has received considerable attention due to their low-power and CMOS-compatible scaling.

Synaptic weights are represented as conductance of the insulating layer within each OxRRAM cell. To program an OxRRAM cell, elevated voltages are applied at the top and bottom electrodes, which re-arranges the atomic structure of the insulating layer. Figure~\ref{fig:RRAM} shows the High-Resistance State (HRS) and the Low-Resistance State (LRS) of an OxRRAM cell. An OxRRAM cell can also be programmed into intermediate low-resistance states, allowing its multilevel operations. 

\begin{figure}[h!]
	\centering
	\vspace{-5pt}
	\centerline{\includegraphics[width=0.89\columnwidth]{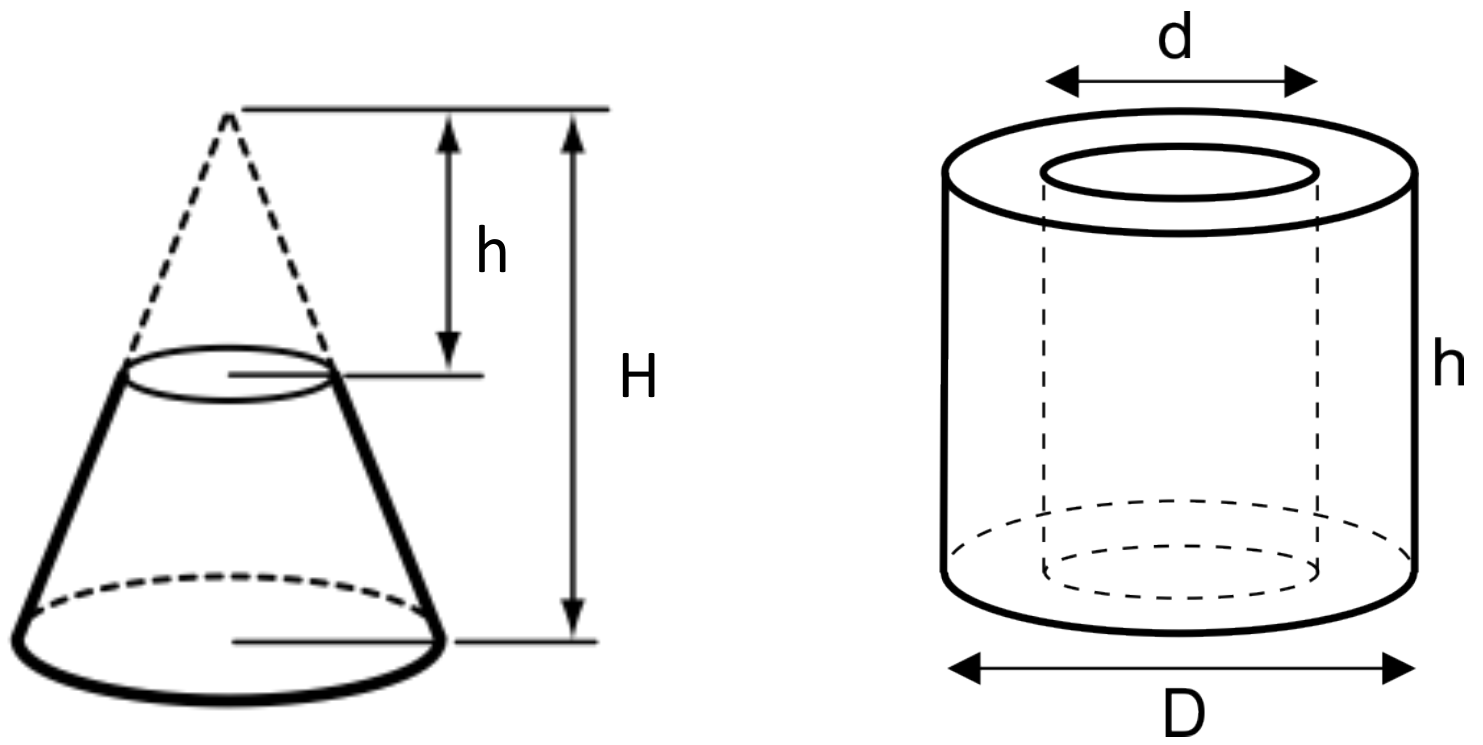}}
	\vspace{-10pt}
	\caption{Read disturbances due to structural alteration in an OxRRAM cell. The left subfigure shows a reduction of the conductive filament gap on application of a stress voltage. Such changes leads to a reduction in the resistance, i.e., a read disturbance of HRS state in the cell. The right subfigure shows the lateral growth of the conductive filament due to application of a stress voltage. This illustrates read disturbance of LRS state in the cell.}
	\vspace{-5pt}
	\label{fig:read_disturbances}
\end{figure}

In OxRRAM technology, the transition from HRS state is governed by a sudden decrease of the vertical filament gap on application of stress voltage during spike propagation~\cite{shim2020impact}.
This is illustrated in the left subfigure of Figure~\ref{fig:read_disturbances} where the vertical filament gap is shown to reduce by an amount $h$. This may result in a conducting filament between the two metal layers causing the resistive state to change from HRS to LRS.
The rate of change of the filament gap of the OxRRAM cell at the \ineq{(i,j)^\text{th}} location in the crossbar is
\begin{equation}
    \label{eq:hrs}
    \footnotesize \frac{dg_{i,j}}{dt} = -\vartheta_0\cdot e^{-\frac{E_a}{kT}}sinh\left(\frac{\gamma_{i,j}\cdot a_0}{L}\cdot\frac{qV_{i,j}}{kT}\right) \text{, where } \gamma_{i,j} = \gamma_0 - \beta\cdot\frac{g_{i,j}}{g_0}^3
\end{equation}
In the above equation, \ineq{t} defines the state transition time, \ineq{g_0} is the initial filament gap of the OxRRAM cell, \ineq{V_{i,j}} is the voltage applied to the cell, \ineq{\gamma_{i,j}} is the local field enhancement factor and is related to the gap \ineq{g_{i,j}}, \ineq{a_0} is the atomic hoping distance, and \ineq{\gamma_0} is a fitting constant.

The transition from one of the LRS states is governed by the lateral filament growth~\cite{shim2020impact}. 
This is illustrated in the right subfigure of Figure~\ref{fig:read_disturbances}.
The time for state transition in the \ineq{(i,j)^\text{th}} RRAM cell is given by 
\begin{equation}
    \label{eq:lrs}
    \footnotesize t_{i,j} (LRS) = 10^{-14.7\cdot V_{i,j} + 6.7} \text{sec}
\end{equation}


If the state transition time of an OxRRAM cell is 1000 ms, then a single quasi-static read operating using 1000 ms read pulse or 1000 read accesses using with 1-ms spike pulses can lead to an abrupt change in the state of the cell. Therefore, the endurance of the cell is 1000. If the cell propagates \ineq{n} spikes during inference of each image, then the inference lifetime is defined as \ineq{1000 / n}. Formally,
\begin{equation}
    \label{eq:inference_lifetime}
    \footnotesize \text{inference lifetime} = \frac{\text{read endurance}}{\text{spikes per image}}
\end{equation}



%% file: sections/dtco.tex
The computer memory industry has thus far been primarily driven by the {cost-per-bit metric}, which provides the maximum capacity for a given manufacturing cost. As shown in recent works~\cite{mutlu2015research,datacon,lee2013tiered,palp,mutlu2013memory,mneme,hebe}, manufacturing cost can be estimated from the area overhead. To estimate the cost-per-bit of a neuromorphic core, we investigate the internal architecture of a crossbar and find that a neuron circuit can be designed using 20 transistors and a capacitor~\cite{indiveri2003low}, while an NVM cell is a 1T-1R arrangement with a transistor used as an access device for the cell. Within an $N \times N$ crossbar, there are \ineq{N} pre-synaptic neurons, \ineq{N} post-synaptic neurons, and \ineq{N^2} synaptic cells. The total area of all the neurons and synapses of a crossbar is

\begin{figure}[h!]
	\centering
	\vspace{-5pt}
	\centerline{\includegraphics[width=0.99\columnwidth]{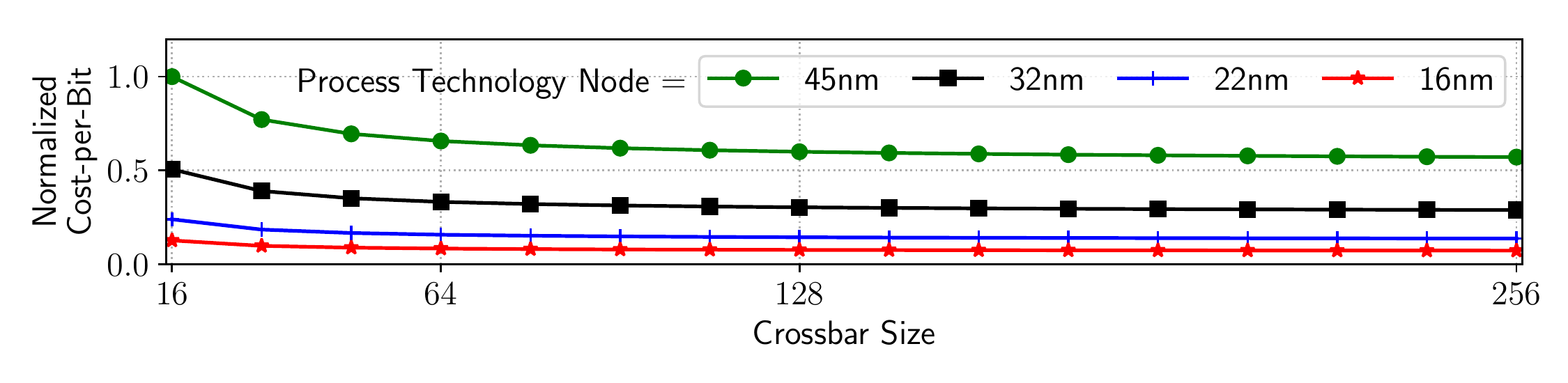}}
	\vspace{-10pt}
	\caption{Cost-per-bit analysis of a crossbar.}
	\vspace{-5pt}
	\label{fig:cost_per_bit}
\end{figure}

\begin{footnotesize}
\begin{eqnarray}
    \label{eq:neuron_synapse_area}
    &\text{neuron area} &= 2N(20T + 1C)\\ \nonumber
    &\text{synapse area} &= N^2(1T + 1R)
\end{eqnarray}
\end{footnotesize}\normalsize
where \ineq{T} stands for transistor, \ineq{C} for capacitor, and \ineq{R} for NVM cell. The total synaptic cell capacity is \ineq{N^2}, with each NVM cell implementing 2-bit per synapse. The total number of bits (i.e., synaptic capacity) in the crossbar is
\begin{equation}
    \label{eq:total_bits}
    \footnotesize \text{total bits} = 2N^2
\end{equation}
Therefore, the cost-per-bit of an \ineq{N}x\ineq{N} crossbar is
\begin{equation}
    \label{eq:cost_per_bit}
    \footnotesize \text{cost-per-bit} = \frac{2N(20T+1C)+N^2(1T+1R)}{2N^2} \approx \frac{F^2(27+2N)}{N},
\end{equation}
where the cost-per-bit is represented in terms of the crossbar dimension \ineq{N} and the feature size \ineq{F}. Equation~\ref{eq:cost_per_bit} provides a {back-of-the-envelope} calculation of cost-per-bit.
Figure~\ref{fig:cost_per_bit} plots the normalized cost-per-bit for four different process technology nodes, with the crossbar dimension ranging from 16 to 256. We observe that the cost-per-bit reduces with increase in the dimension of a crossbar, i.e., larger-sized crossbars can accommodate more bits for a given cost.

\begin{figure}[h!]%
    \centering
    \subfloat[Current map in a 128x128 crossbar.\label{fig:current_map}]{{\includegraphics[width=4.0cm]{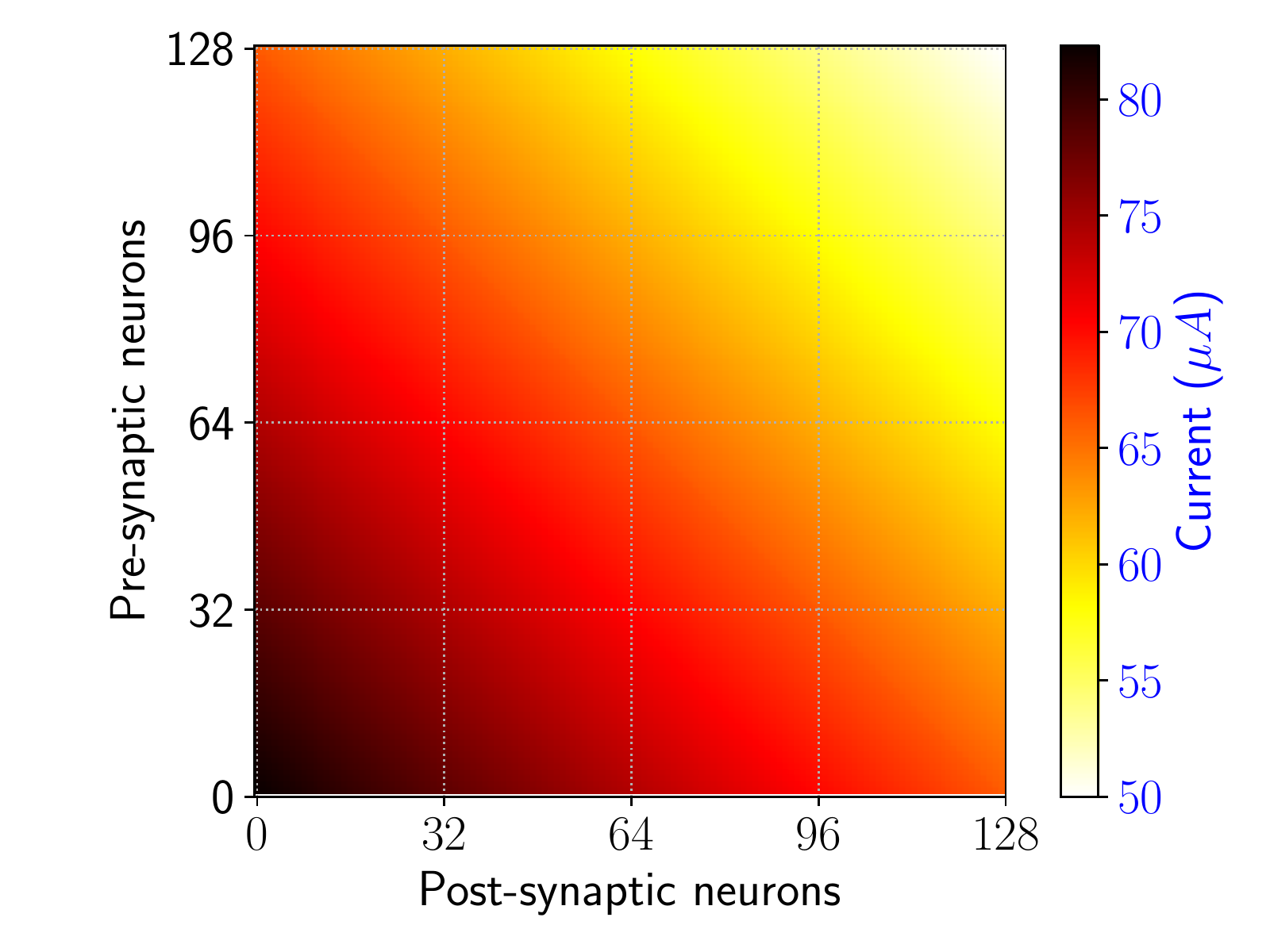} }}%
    \quad
    \subfloat[RRAM endurance in a  128x128 crossbar.\label{fig:hrs_var}]{{\includegraphics[width=4.0cm]{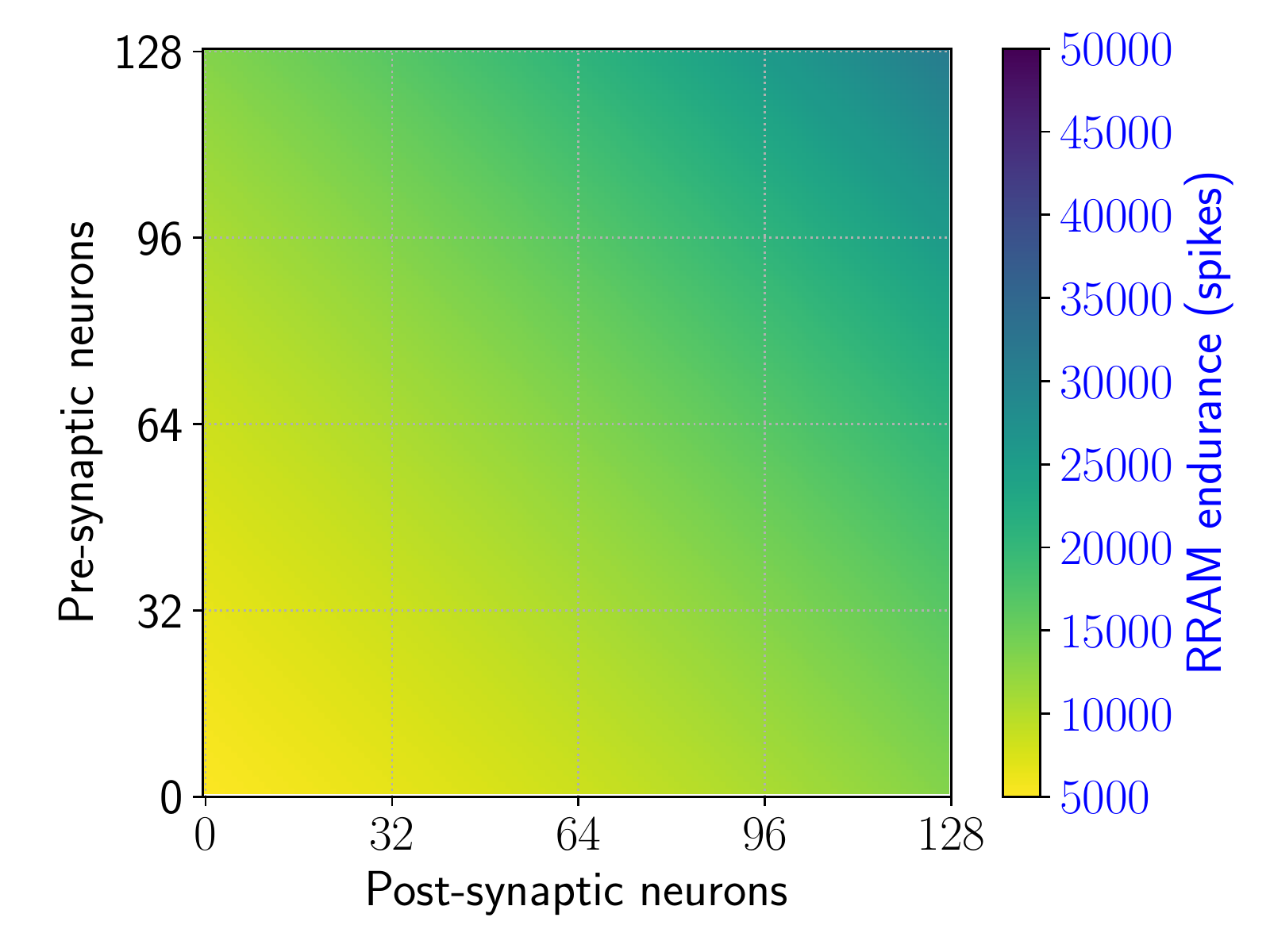} }}%
    \caption{Current map and RRAM endurance in a  128x128 crossbar.}%
    \label{fig:current_endurance_map}%
\end{figure}

We now analyze the internal architecture of a crossbar. 
Figure~\ref{fig:current_map} shows the current through the memory cells in a 128x128 crossbar. This current variation is due to the difference in the length of current paths from pre to post-synaptic neurons in the crossbar, where the length of a current path is measured in terms of the number of parasitic components on the path. These current values are obtained for a 65nm technology node and at 300K temperature corner. As can be clearly seen from the figure, current through memory cells on the top-right corner of the crossbar is lower than those at the bottom-left corner.


Figure~\ref{fig:hrs_var} shows the endurance variation of a 128x128 crossbar at 45~nm node and at 30$^\circ$C with each RRAM cell programmed to HRS state. The endurance variation is a direct result of the current variation in the crossbar.

\begin{figure}[h!]
	\centering
	\centerline{\includegraphics[width=0.99\columnwidth]{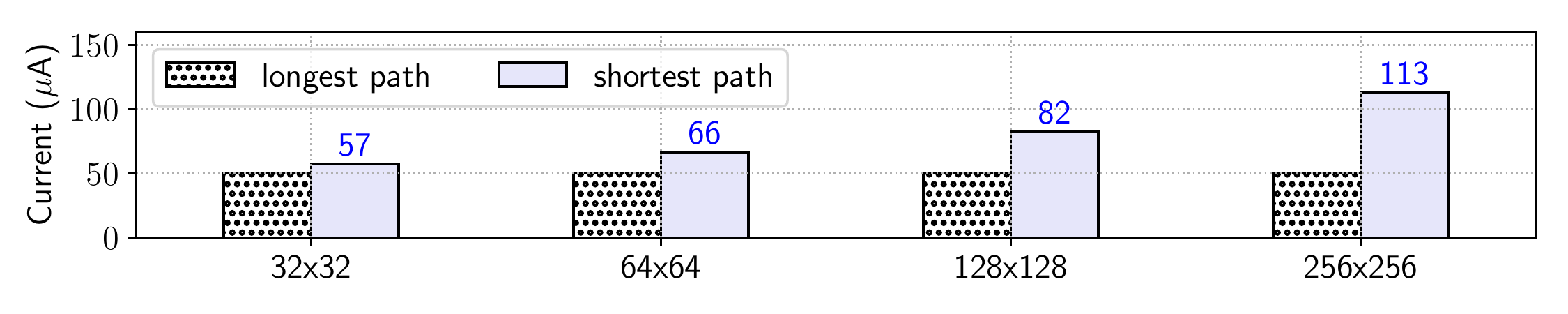}}
	\caption{Difference between current on the shortest and the longest path for different crossbar sizes.}
	\label{fig:current_crossbar_size}
\end{figure}

Figure~\ref{fig:current_crossbar_size} shows the difference between currents on the shortest and longest paths for 32x32, 64x64, 128x128, and 256x256 crossbars at {65nm} process node. The input spike voltage of the pre-synaptic neurons is set to generate \ineq{50\mu A} on the longest path. This current value corresponds to the current needed to read the resistance state of an OxRRAM cell. 

We observe that the current injected into the post-synaptic neuron on the longest path is lower than the current on the shortest path by 13.3\% for 32x32, 25.1\% for 64x64, 39.2\% for 128x128, and 55.8\% for 256x256 crossbar.
This current difference is because of the higher voltage drop on the longest path, which reduces the current on this path compared to the shortest path for the same amount of spike voltage applied on both these paths.
The current difference increases with crossbar size because of the increase in the number of parasitic resistances on the longest current path, which results in larger voltage drops, lowering the current injected into its post-synaptic neuron. However, larger current variation causes larger endurance variation as illustrated in Figure~\ref{fig:hrs_var}.
Therefore, larger crossbar sizes leads to larger endurance variation.

System designers often make a tradeoff between cost-per-bit and endurance variation. Typically, crossbar sizes of \ineq{128\times128} and \ineq{256\times256} gives the best tradeoff.

%% file: sections/df.tex
An optimized design flow is one, which takes into account the design and technology characteristics to place neurons and synapses to a crossbar such that those synapses that propagate more spikes are mapped to NVMs with higher read endurance. This is to increase the inference lifetime.

\begin{figure}[h!]
	\begin{center}
		\includegraphics[width=0.99\columnwidth]{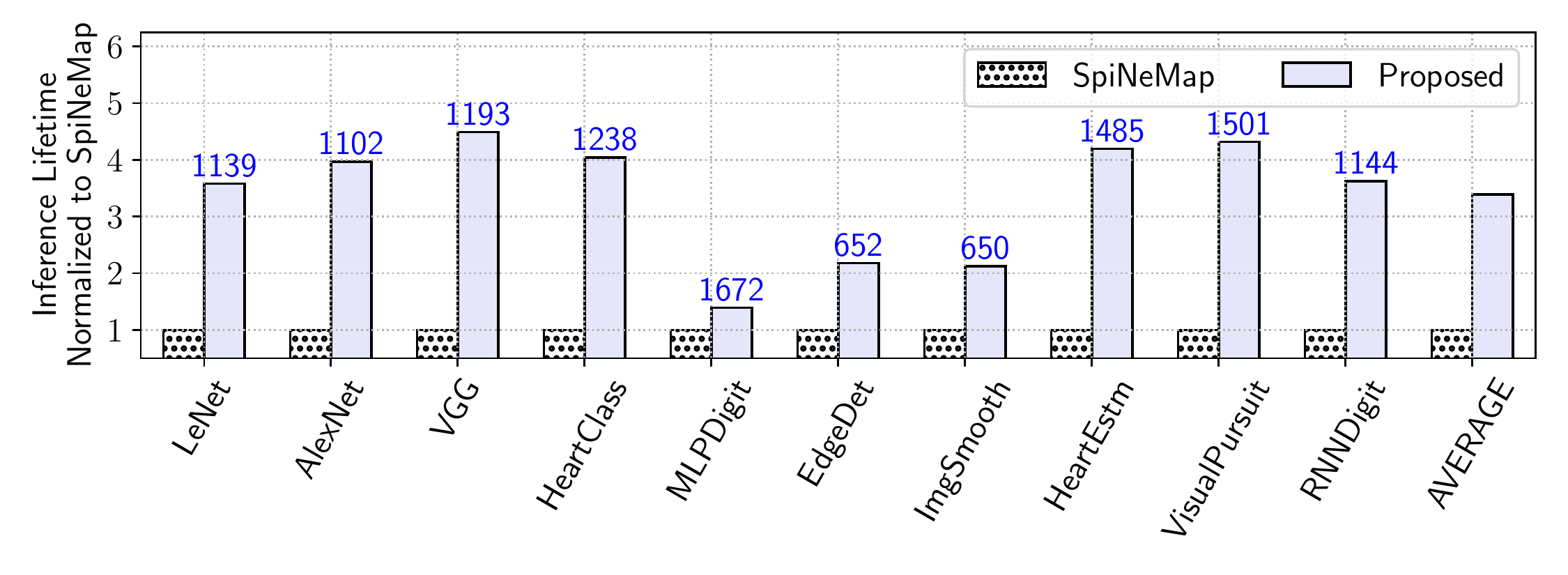}
		\caption{Inference lifetime normalized to \sm{}.}
		\label{fig:il_unlimited}
	\end{center}
\end{figure}

In our prior work~\cite{song2021improving}, we show that such a design flow can significantly improve the inference lifetime.
Figure~\ref{fig:il_unlimited} reports the inference lifetime for 10 applications for the proposed design flow normalized to \sm{}. We observe that through intelligent synapse mapping, the inference lifetime obtained using the proposed approach is on average 3.4x higher.

%% file: sections/conclusions.tex

A design-technology tradeoff analysis is performed to investigate inference lifetime of neuromorphic hardware that adopts NVM as synaptic storage. An essential observation of the analysis is that the read endurance of an NVM cell depends on its programmed synaptic weight (design) and the voltage exposed to the cell (technology). Our analysis also reveals that voltages exposed to NVM cells inside a crossbar vary due to the parasitic components on current paths, which leads to asymmetric read endurance across NVM cells in a crossbar. From detailed circuit-level simulations, we show design scaling on read endurance of NVMs negatively impacts the inference lifetime of neuromorphic hardware. In addition, the design flow that considers asymmetry in read endurance can significantly improve the inference lifetime of neuromorphic hardware.